\documentclass[12pt]{article}

\begin{document}

\title{Inevitability of Spacetime Singularities in The Canonical Metric}
\author{J. Ponce de Leon\thanks{E-mail: jponce@upracd.upr.clu.edu or jpdel@coqui.net}\\ Laboratory of Theoretical Physics, Department of Physics\\ 
University of Puerto Rico, P.O. Box 23343, San Juan, \\ PR 00931, USA}
\date{June 2001}

\maketitle
\begin{abstract}
We discuss the question of whether the existence of singularities is an intrinsic property of $4D$ spacetime. Our hypothesis is that singularities in $4D$ are induced by the separation of spacetime from the other dimensions. We examine this hypothesis in the context of the so-called canonical or warp metrics in $5D$. These metrics are popular because they provide a clean separation between the extra dimension and spacetime. We show that the spacetime section, in these metrics, inevitably becomes singular for some finite (non-zero) value of the extra coordinate. This is true for all canonical metrics that are solutions of the field equations in space-time-matter theory. This is a geometrical singularity in $5D$, but appears as a physical one in $4D$. At this singular hypersurface, the determinant of the spacetime metric becomes zero and the curvature of the spacetime blows up to infinity. These results are consistent with our hypothesis.

 \end{abstract}



\newpage

\section{Introduction} 
An important question in gravitational theories is whether the singularities that appear in the theory are intrinsic properties of the spacetime. This is especially relevant to models of our universe formulated in more than four-dimensions. Even if the model is free of singularities in $D$-dimensions, its interpretation in one lower dimension might induce singularities.

A well known example that illustrates this point is provided by the metric \cite {JPdeL 1} 
\begin{equation}
\label{Ponce de Leon solution}
d{\cal S}^2 = (x^4)^2 dt^2 - t^{2/\alpha}(x^4)^{2/(1 - \alpha)}[dr^2 + r^2(d\theta^2 + \sin^2\theta d\phi^2)] - \alpha^2(1- \alpha)^{-2} t^2 (dx^4)^2.
\end{equation}
Here $\alpha$ is a constant, $x^4$ is the coordinate along the  extra-dimension and $t, r, \theta$ and $\phi$ are the usual coordinates for a spacetime with spherically symmetric spatial sections. 

The properties of this metric have been discussed in many places \cite{Overduin}-\cite{Wesson hd}. This is a solution to the field equations ${\cal R}_{AB} = 0$, in five dimensions. It is free of singularities of any kind in $5D$. In four-dimensions (on the hypersurfaces $x^4 = Constant$) this metric corresponds to the $4D$ Friedmann-Robertson-Walker models with flat $3D$ sections. The equation of state of the effective perfect fluid in $4D$ is: $p = n \rho$ with $n = (2\alpha/3 -1)$ ($\alpha = 2$ for radiation, $\alpha = 3/2$ for dust, etc.). Thus, although (\ref{Ponce de Leon solution}) is free of singularities in $5D$, our interpretation in $4D$ induces a big bang singularity. 

The above example leads us to the hypothesis that singularities in $4D$ are not a property of the spacetime, but a result of the separation between the extra coordinate and spacetime. Here, we will study this hypothesis using the so-called canonical or warp type metric in $5D$. 

These metrics provide a clear separation between the extra coordinate and the spacetime section. We will show that, as a result of the field equations in $5D$, the embedded $4D$ spacetime becomes singular for some finite value of the extra coordinate. 

This means that the ``splitting" in $(1 + 4)$, provided by the canonical metrics, induces singularities in the physical $4D$ spacetime.
\section{Canonical or Warp Metric}

In order to set our notation, let us start by providing some general background on the canonical metric\footnote{For more detailed discussion we refer the reader to Refs. \cite{Wesson book} and \cite{Wesson hd}.} This metric is useful not only in five-dimensions, but in any number of dimensions.  

Consider a $D$-dimensional ``generating" space that we foliate by a family of $(D - 1)$ hypersurfaces. In Gaussian normal coordinates the metric of the generating space reads 
 
\begin{equation}
\label{Gauss metric}
d{\cal S}^2 = \gamma_{AB}dx^Adx^B = \gamma_{\mu\nu}(x^{\rho}, y) dx^{\mu}dx^{\nu}  + \epsilon dy^2,
\end{equation}
where $A,B = 0,1...(D - 1)$; $\mu, \nu = 0,1..(D - 2)$. The factor $\epsilon$ is taken to be $+1$ or $-1$ depending on whether the hypersurfaces are timelike or spacelike. The physical metric in $(D -1)$ is conformally related to the induced one, viz.,
\begin{equation}
\label{Warp metric}
d{\cal S}^2 =  \Omega g_{\mu\nu}(x^{\rho}, y)dx^{\mu}dx^{\nu} + \epsilon dy^2,
\end{equation}
The conformal factor $\Omega$ is called {\it warp} factor, and 
 $g_{\mu\nu}(x^\rho, y)$ is interpreted as the {\it physical} metric on the embedded hypersurface of one lower dimension. 

The factor $\Omega$ can be a function of $x^{\mu}$ and $y$, in principle. In ``brane worlds" scenarios it is taken in the form $\Omega = e^{-ky}$ or $\Omega = (1 - ay)^{\alpha}$, where the coefficients $k$, $a$ and $\alpha$ depend, among other considerations, on the number of dimensions \cite{Youm}-\cite{Deruelle and Katz}.   
 
In space-time-matter theory, where $D = 5$, the factor $\Omega$ is taken as $\Omega = (y/L)^2$, where $L$ is a constant length, which is later identified in terms of the cosmological constant via $\Lambda = 3/L^2$. Namely,
\begin{equation}
\label{canonical metric}
d{\cal S}^2 = \frac{y^2}{L^2}g_{\mu\nu}(x^{\rho}, y)dx^{\mu}dx^{\nu} + \epsilon dy^2,
\end{equation}
This metric is usually called {\it canonical} metric. This and (\ref{Warp metric}) have been discussed in many different contexts and applied to different problems. In particular, to the motion of test particles and the appearance of an extra force acting on the particle, as measured by an observer in one lower dimension.

\section{Field Equations in Canonical Coordinates}

For the canonical line element (\ref{canonical metric}), the components of the Ricci tensor, in terms of the induced $4D$ metric $\hat{g}_{\mu\nu} = \Omega(y)g_{\mu\nu}$, are given by\footnote{For the signature of the metric we use $(+ - - - {\epsilon})$. } \cite{last work} 
\begin{eqnarray}
\label{Ricci Tensor}
\hat{\cal R}_{(4)(4)}&=& - (\hat{\Psi}_{|(4)} + \hat{\Psi}_{\lambda\rho}\hat{\Psi}^{\lambda\rho}),\nonumber \\
\hat{\cal R}_{(\mu)(4)} &=& \hat{D}_{(\lambda)}(\hat{\Psi}_{\mu}^{\lambda} - \delta_{\mu}^{\lambda}\hat{\Psi}),\nonumber \\
\hat{\cal R}_{(\mu)(\nu)} &=& \hat{R}_{\mu\nu} - \epsilon(\hat{\Psi}_{\mu\nu|(4)} - 2 \hat{\Psi}_{\mu\rho}\hat{\Psi}^{\rho}_{\nu} + \hat{\Psi}\hat{\Psi}_{\mu\nu}).
\end{eqnarray}
Here ``$|(4)$" denotes $\epsilon (\partial/\partial y)$ and\footnote{All ``hated" quantities are calculated with the induced metric $\hat{g}_{\mu\nu}= \Omega g_{\mu\nu}$.}    
\begin{equation}
\hat{\Psi}_{\mu\nu} = \frac{1}{2}\hat{g}_{\mu\nu|(4)},\;\; \hat{\Psi}^{\mu\nu} = -\frac{1}{2}(\hat{g}^{\mu\nu})_{|(4)}, \;\; \hat{\Psi}^{\alpha}_{\beta} = \hat{g}^{\alpha\lambda}\hat{\Psi}_{\lambda\beta}.
\end{equation}
In our discussion we will need only the first equation in (\ref{Ricci Tensor}). 
 For an arbitrary warp factor, it reads\footnote{$( )^{\ast}$ indicates differentiation with respect to $y$, viz., $\partial()/ \partial y = ( )^{\ast}$.}
\begin{equation}
\label{R44, arbitrary warp function}
\hat{\cal R}_{(4)(4)} = - \left[\frac{1}{2}(g^{\mu\nu}\stackrel{\ast}{g}_{\mu\nu})^{\ast} + \frac{\stackrel{\ast}{\Omega}}{ 2\Omega}(g^{\mu\nu}\stackrel{\ast}{g}_{\mu\nu}) - \frac{1}{4}\stackrel{\ast}{g}_{\mu\nu}(g^{\mu\nu})^{\ast} + 2 \frac{\stackrel{\ast \ast}{\Omega}}{\Omega} - \frac{(\stackrel{\ast}{\Omega})^2}{\Omega^2}\right].
\end{equation}

Now, setting $\Omega= (y^2/L^2)$, we obtain
\begin{equation}
\label{R44 in explicit form}
\hat{\cal R}_{(4)(4)} = -\left[\frac{1}{2}(\stackrel{\ast}{g}_{\mu\nu}g^{\mu\nu})^{\ast} + \frac{1}{y}g^{\mu\nu}\stackrel{\ast}{g}_{\mu\nu} - \frac{1}{4}\stackrel{\ast}{g}_{\mu\nu}(g^{\mu\nu})^{\ast}\right].
\end{equation}
We should require $\Omega \neq 0$ in (\ref{R44, arbitrary warp function}) and $y \neq 0$ in (\ref{R44 in explicit form}). We now use the field equations in $5D$ ($\hat{\cal R}_{AB} = 0$, see Ref. \cite{Wesson book}) and make the change of variable
\[
z =  \frac{C}{y},\;\;\ C = Constant,\;\; y \neq 0.
\]
Substituting into (\ref{R44 in explicit form}) we obtain
\begin{equation}
\label{R44 in terms of z}
(g_{\mu\nu,z}g^{\mu\nu})_{,z} - \frac{1}{2}g_{\mu\nu,z}(g^{\mu\nu})_{,z} = 0.
\end{equation}
We now introduce the notation 
\begin{equation}
\label{notation}
\chi_{\mu\nu} = g_{\mu\nu,z} = \frac{\partial g_{\mu\nu}}{\partial z},\;\;\chi_{\nu}^{\lambda} = g^{\mu\lambda}\chi_{\mu\nu}.
\end{equation}
This quantity is a four-dimensional tensor. Its trace, $\chi^{\alpha}_{\alpha}$, can be expressed as 
\begin{equation}
\label{log of metric}
\chi^{\alpha}_{\alpha} = \frac{1}{g}\frac{\partial g}{\partial z} = \frac{\partial}{\partial z}\ln(-g).
\end{equation}
Since $(g^{\mu\lambda})_{,z} = - g^{\mu\nu}\chi^{\lambda}_{\nu}$, Eq. (\ref{R44 in terms of z}) becomes
\begin{equation}
\label{resembles synchronous coordinates} 
\frac{\partial}{\partial z}\chi^{\alpha}_{\alpha} + \frac{1}{2}\chi_{\mu}^{\nu}\chi^{\mu}_{\nu} = 0.
\end{equation}
This equation resembles the one in the synchronous system of reference as discussed in Ref. \cite{Landau and Lifshitz}. 
\subsection{Inevitability of Singularities}
We now proceed to show that the spacetime sections of the canonical metric become singular for some finite value of the extra coordinate. 

Using the inequality
\begin{equation}
\label{inequality}
\chi_{\alpha}^{\beta}\chi^{\alpha}_{\beta}\geq \frac{1}{4}(\chi^{\alpha}_{\alpha})^2
\end{equation}
in (\ref{resembles synchronous coordinates}) we obtain
\begin{equation}
\frac{\partial}{\partial z}\chi^{\alpha}_{\alpha} + \frac{1}{8}(\chi^{\alpha}_{\alpha})^2\leq 0.
\end{equation}
From which it follows that 
\begin{equation}
\label{basic equation}
\frac{\partial}{\partial z}\left(\frac{1}{\chi^{\alpha}_{\alpha}}\right) \geq \frac{1}{8}.
\end{equation}
This equation constitutes the basis for our discussion. It shows that $(1/\chi^{\alpha}_{\alpha})$ is a monotone function of $z$. Thus, if  $\chi^{\alpha}_{\alpha} > 0$, for some value of $z$, then the function $(1/\chi^{\alpha}_{\alpha})$ will be monotone decreasing with the decrease of $z$. Therefore it will become zero (from the positive side) for some finite value of $z = z_{0}$ ($y \neq 0$). This means, $\chi^{\alpha}_{\alpha}$ becomes $ + \infty$ for that value. Equation (\ref{log of metric}) then shows that the determinant of the metric vanishes at that point. If $\chi^{\alpha}_{\alpha} < 0$, for some value of $z$, the same behavior will occur for increasing $z$. In both cases $- g \sim |z - z_{0}|^a$, where $0 < a \leq 8$. 

The scalar curvature, of the physical spacetime, is given by
\begin{equation}
\label{scalar curvature with Omega}
R = g^{\mu\nu}R_{\mu\nu} = \epsilon \Omega\left[3\left(\frac{\stackrel{\ast}{\Omega}}{\Omega}\right)^2 + \frac{3}{2}\left(\frac{\stackrel{\ast}{\Omega}}{\Omega}\right) \stackrel{\ast}{g}_{\mu\nu}g^{\mu\nu} + \frac{1}{4}\left((\stackrel{\ast}{g}_{\mu\nu}g^{\mu\nu})^2 + \stackrel{\ast}{g}_{\mu\nu}({g}^{\mu\nu})^{\ast}\right)\right].      
\end{equation} 
Setting $\Omega = (y/L)^2$, we get
\begin{equation}
R = \frac{\epsilon}{L^2}\left[ 12 - 3 z \chi^{\alpha}_{\alpha} + \frac{1}{4}z^2((\chi^{\alpha}_{\alpha})^2 - \chi_{\alpha}^{\beta}\chi^{\alpha}_{\beta})\right].
\end{equation}
Thus, $R$ diverges as $R \sim 1/(z - z_{0})^{2}$.

The properties of the induced matter, as calculated by an observer in $4D$, are given by the energy momentum tensor\footnote{See Ref. \cite{last work}, Eq. (62). } (in terms of $z$)
\begin{equation}
T_{\mu\nu} =  -\frac{3\epsilon}{L^2}g_{\mu\nu} + \frac{\epsilon}{2L^2}(\chi_{\mu\nu} - \frac{1}{2}\chi_{\alpha}^{\alpha}g_{\mu\nu})(\frac{z^2}{2}\chi^{\alpha}_{\alpha} - 4z) + \frac{\epsilon}{L^2}z\chi_{\mu\nu} - \frac{\epsilon z^2}{2L^2}(\chi_{\mu\nu,z} + \chi_{\mu\rho}\chi_{\nu}^{\rho} - \frac{1}{4}\chi_{\alpha}^{\beta}\chi_{\beta}^{\alpha}g_{\mu\nu})
\end{equation}
Since $\chi_{\alpha}^{\alpha}$ and $\chi_{\alpha}^{\beta}\chi_{\beta}^{\alpha}$ diverge at $z = z_{0}$, it follows that some terms in the above expression will also diverge.  This completes the demonstration that the physical spacetime, with metric $g_{\mu\nu}$, will inevitably become singular for some finite value of $y$.

A particular solution with these properties is discussed in Ref. \cite{Wesson book}(page 121). That solution is {\it free} of singularities in five-dimensions. But its four-dimensional interpretation exhibits the kind of singularities discussed above. Again the kind of behavior we noted in the Introduction. 
 
\subsection{Arbitrary Warp Factor}

The question may arise of whether the above singularity is not a consequence of the particular choice of the warp factor, $\Omega(y) = y^2/L^2$, used in (\ref{canonical metric}). 

First notice that the above discussion excludes the value $y = 0$ from consideration. This is because this is a singularity in $5D$. All equations diverge for $y = 0$ ($z = \infty$).

Similarly, for an arbitrary warp factor $\Omega(y)$, the generating $5D$ space becomes singular at the hypersurfaces where $\Omega(y) = 0$. This can be seen from (\ref{R44, arbitrary warp function}). Therefore, we should concentrate on finite positive $\Omega(y)$, for which the coordinate system and field equations in $5D$ are well defined.

The whole discussion can be repeated in terms of the induced metric $\hat{g}_{\mu\nu} = \Omega(y)g_{\mu\nu}$. In this case, instead of (\ref{basic equation}) we obtain 
\begin{equation}
\frac{\partial}{\partial y}\left(\frac{1}{\hat{\chi}^{\alpha}_{\alpha}}\right) \geq \frac{1}{8},
\end{equation} 
with $\hat{\chi}_{\alpha\beta} = \partial(\hat{g}_{\alpha\beta})/\partial y$. Therefore, we arrive at the same conclusions as before. In particular, the curvature 
\[
\hat {R} = \hat{g}^{\mu\nu}\hat{R}_{\mu\nu} = \frac{\epsilon}{4}\left[(\hat{\chi}^{\alpha}_{\alpha})^2 - \hat{\chi}_{\alpha}^{\beta}\hat{\chi}_{\beta}^{\alpha}\right],
\]
by virtue of the inequality (\ref{inequality}), inevitably diverges for some finite value of $y$ (where $\Omega(y) \neq 0$). The scalar curvature $R$ of the physical spacetime is given by
\begin{equation}
R = \Omega\hat{R} + 3 \left( \frac{\Omega^{\mu}_{;\mu}}{\Omega} - \frac{1}{2}\frac{\Omega^{\mu}\Omega_{\mu}}{\Omega^2}   \right). 
\end{equation}
In our case this reduces to $R = \Omega \hat{R}$. Since $\Omega \neq 0$, it follows that $R$ diverges where  $\hat{R}$ does it.

Consequently, the singularity in the spacetime section of the metric in the canonical form is independent of the particular choice of the warp factor. 
  
\section{Discussion and Conclusions}
In canonical coordinates, the $y$-lines ($x^{\mu} = Const.$) are geodesics lines orthogonal to $4D$ spacetime. Therefore, in these coordinates there is a clear separation between the extra coordinate, valid everywhere in spacetime, and the metrical description of spacetime itself. 

The vanishing of the determinant of the spacetime metric on the hypersurface $z = z_{0}$ indicates that the separation between these geodesics becomes zero at the point of intersection with this hypersurface. This means that the singularity is one of geometrical nature (in $5D$) associated with the specific properties of the coordinates used. Therefore, this singularity is not a genuine one (in $5D$) and, {\it can be eliminated} by making the appropriate change to another (non-canonical) system of coordinates in $5D$.

The above discussion is similar to the one about singularities in the synchronous frame of reference, as provided in Landau and Lifshits \cite{Landau and Lifshitz}. 

However, to an observer ``living" in $4D$ this induced singularity will look as one of  physical nature. Firstly because of the blowing up of the spacetime curvature, which he will interpret, in terms of the properties of matter, as diverging energy density and pressure. 

Perfect examples that illustrate this point are provided by the cosmological solution discussed in the Introduction and the ``shell-like" solution discussed in Ref. \cite{Wesson book}, on page 117.  

In conclusion, the nice separation between the extra coordinate and the $4D$ spacetime, in canonical coordinates, is not free of cost. The price we pay for it is the resulting singularity in $4D$. This result is independent on whether the extra dimension is spacelike or timelike.

\end{document}